\documentclass[10pt]{blois07}
\usepackage{wrapfig,psfrag,epsfig,graphicx,graphics,xcolor}
\usepackage[small]{caption2}
\usepackage{amssymb}
\usepackage{amsmath}
\usepackage{amsfonts}
\usepackage{cite,./mcite}

\def\lsim{\raise0.3ex\hbox{$<$\kern-0.75em\raise-1.1ex\hbox{$\sim$}}}
\def\gsim{\raise0.3ex\hbox{$>$\kern-0.75em\raise-1.1ex\hbox{$\sim$}}}
\def\noi{\noindent}

\def\bei{\begin{itemize}}
\def\ei{\end{itemize}}
\def\bea{\begin{eqnarray}}
\def\eea{\end{eqnarray}}
\def\beas{\begin{eqnarray*}}
\def\eeas{\end{eqnarray*}}
\def\beqas{\begin{eqnarray*}}
\def\eqas{\end{eqnarray*}}
\def\beq{\begin{equation}}
\def\eq{\end{equation}}
\def\eeq{\end{equation}}
\def\beqd{\begin{displaymath}}
\def\eeqd{\end{displaymath}}
\def\eqd{\end{displaymath}}

\def\beeq{\begin{eqnarray}} \def\eeeq{\end{eqnarray}}

\def\bef{\begin{frame}}

\def\slashchar#1{\setbox0=\hbox{$#1$}
   \dimen0=\wd0
   \setbox1=\hbox{/} \dimen1=\wd1
   \ifdim\dimen0>\dimen1
      \rlap{\hbox to \dimen0{\hfil/\hfil}}
      #1
   \else
      \rlap{\hbox to \dimen1{\hfil$#1$\hfil}}
      /
   \fi}

\newcommand{\widttt}{0.185\columnwidth}

\newcommand{\rb}{\underline{r}}
\newcommand{\kb}{\underline{k}}

\newcommand{\fin}{\end{document}}
\newcommand{\vn}{\vspace{-.5cm}\noindent}

\usepackage[english]{babel}

\usepackage[latin1]{inputenc}

\usepackage{times}
\usepackage[T1]{fontenc}

\begin{document}
\title{Perturbative QCD in the Regge limit: Prospects at ILC}
\author{S.~Wallon$^1$\protect\footnote{ \, talk presented at EDS07}}
\institute{$^1$\,LPT, Universit\'e
Paris-Sud, CNRS, Orsay, France}
\maketitle

\begin{abstract}
After recalling the theoretical and experimental status of QCD in the Regge limit and the requirement of
high energy scattering process of onium-onium type for testing this limit, we show that the International Linear Collider would be a major step in this field. 
\end{abstract}

\section{QCD in the Regge limit: theoretical status}
\subsection{LL BFKL Pomeron}

\begin{wrapfigure}{r}{0.3\columnwidth}
\hspace{-.3cm}
\psfrag{M1}[cc][cc]{\scalebox{.65}{$\hspace{-.5cm}M_1^2 \gg \Lambda_{QCD}^2$}}
\psfrag{M2}[cc][cc]{\scalebox{.65}{$\hspace{-.5cm}M_2^2 \gg \Lambda_{QCD}^2$}}
\psfrag{s}[cc][cc]{\scalebox{.6}{$s \rightarrow$}}
\psfrag{t}[cc][cc]{\scalebox{.6}{$\begin{array}{c}t \\ \downarrow \end{array}$}}
\psfrag{v}[ll][ll]{\hspace{-.5cm}\tiny\begin{tabular}{c} $ \leftarrow $ vacuum quantum  \\ number \end{tabular}}
\psfrag{i1}[ll][ll]{\hspace{-.1cm}\tiny impact factor}
\psfrag{i2}[ll][ll]{\hspace{-.1cm}\tiny impact factor}
\begin{center}\vspace{-.8cm}\hspace{-.5cm}
\raisebox{.2 \totalheight}{\epsfig{file=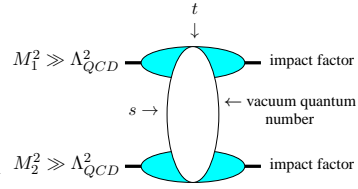,width=2.9cm}} 
\end{center}
\vspace{-.8cm}
\caption{Scattering at $s \gg -t$.}
\vspace{-.5cm}
\label{FigscatteringOnium}
\end{wrapfigure}
\begin{figure}[h]
$$ \begin{array}{ccccccc} \raisebox{-0.45 \totalheight}{\hspace{-.9cm}\epsfig{file=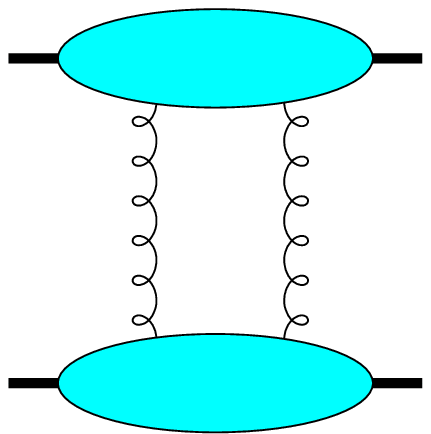,width=1.6cm}} &\hspace{-2.8cm}+\hspace{-2.8cm}&  \left(\,   \raisebox{-0.45 \totalheight}{\epsfig{file=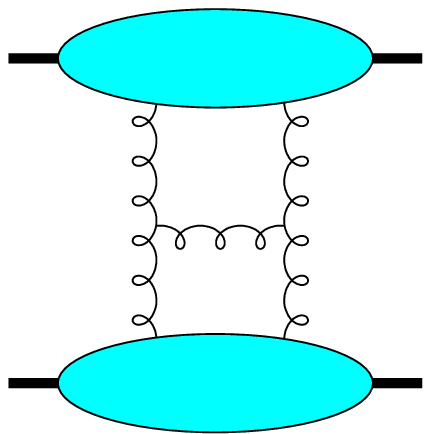,width=1.6cm}} \hspace{0cm}+  \raisebox{-0.45 \totalheight}{\epsfig{file=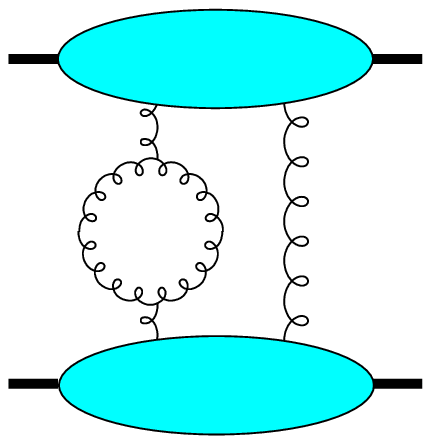,width=1.6cm}} \hspace{-.4cm}+ \cdots \right ) &\hspace{-1.4cm}+\hspace{-1.1cm}& \hspace{-.3cm}\left(\raisebox{-0.45 \totalheight}{\epsfig{file=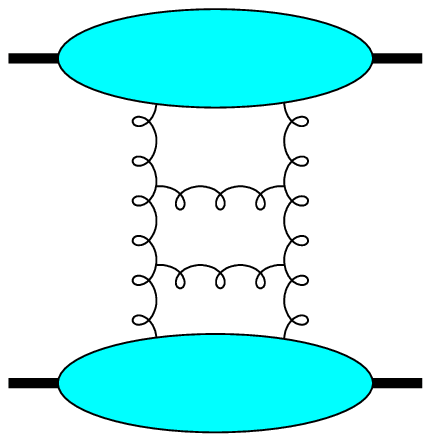,width=1.6cm}}\hspace{-.3cm}+ \cdots \right) + \cdots &\hspace{-.2cm}=\hspace{-.6cm}& \scalebox{.55}{\raisebox{-0.46
    \totalheight}{\epsfig{file=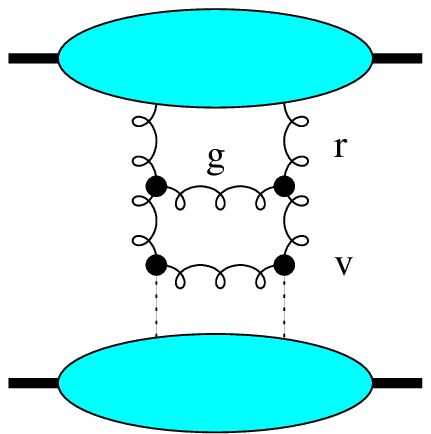,width=3.2cm}}}\\
 \raisebox{.4
    \totalheight}{\hspace{-1cm} $\sim1 $}& & \raisebox{.4
    \totalheight}{\hspace{-1cm}$\sim \alpha_S \ln s$} & &\raisebox{.4
    \totalheight}{\hspace{-1.5cm} $ \sim (\alpha_S \, \ln s)^2 $} 
   \end{array}$$\vspace{-.8cm}
\caption{BFKL resummation.}\vspace{-.65cm}
\label{FigBFKLresum}
\end{figure}

At high energy ($s \gg -t$),
consider the elastic scattering amplitude of two {\it IR safe (hard) probes} (Fig.\ref{FigscatteringOnium}).
 Small values of $\alpha_S$ \linebreak
(perturbation theory applies due to hard scales) 
are compensated by  large
$\ln s,$ 
calling for a resummation of $\sum_n (\alpha_S \, \ln s)^n$ series, resulting in the effective  BFKL ladder \cite{bfkl}, the Leading Log
  hard Pomeron (Fig.\ref{FigBFKLresum}). Optical theorem gives
$\sigma_{tot} \sim s^{\alpha_P(0) -1}$
with 
$ \alpha_P(0)\!-\!1=C \, \alpha_S$. Since $C >0,$
Froissart bound is violated \nolinebreak[4] at \linebreak perturbative order. The large $N_c$ color dipole model\cite{dipoleNik,dipoleMueller}, 
based on perturbation theory on the light-cone, is equivalent to BFKL approach
\psfrag{g}[cc][cc]{\small gluon}
\psfrag{r}[ll][ll]{\small reggeon}
\psfrag{v}[ll][ll]{\small effective vertex}\vspace{-.2cm}
\vspace{-.3cm}
  at the level of diagrams  and   amplitude \cite{MuellerChenNaveletSW}.

\subsection{$k_T$ factorization ($\gamma^* \gamma^* \to \gamma^* \gamma^*$ case)}

\psfrag{l1}[cc][cc]{$$}
\psfrag{l1p}[cc][cc]{$$}
\psfrag{l2}[cc][cc]{$$}
\psfrag{l2p}[cc][cc]{$$}
\psfrag{ai}[cc][cc]{\scalebox{.8}{$ \beta^{\,\nearrow}$}}
\psfrag{bd}[cc][cc]{\scalebox{.8}{$ \alpha_{\,\searrow}$}}
\psfrag{k}[cc][cc]{\scalebox{.8}{$\!k$}}
\psfrag{rmk}[cc][cc]{\scalebox{.8}{$\, \, r-k$}}
\psfrag{oa}[cc][cc]{\raisebox{.9 \totalheight}{\scalebox{.7}{\hspace{-.2cm}${}\quad {  \alpha_k \ll \alpha_{\rm quarks}}$}}}
\psfrag{g1}[ll][ll]{$\gamma^*$}
\psfrag{g2}[ll][ll]{$\gamma^*$}
\psfrag{p1}[ll][ll]{}
\psfrag{ob}[cc][cc]{\raisebox{-1.4 \totalheight}{\scalebox{.7}{\hspace{-.2cm}${} \quad { \beta_k \ll \beta_{\rm quarks}}$}}}
\psfrag{p2}[ll][ll]{}
\begin{wrapfigure}{r}{0.29\columnwidth}\vspace{-1.2cm}
\begin{center}
\hspace{-.2cm}\epsfig{file=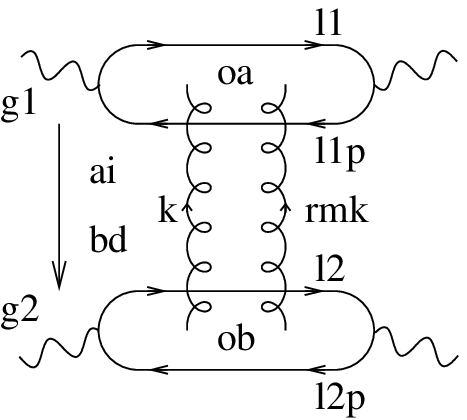,width=0.3\columnwidth}
\end{center}\vspace{-1.2cm}
\caption{$k_T$ factorization.}\vspace{-.4cm}
\label{FigkT}
\end{wrapfigure}
Using the Sudakov decomposition
$k = \alpha p_1 + \beta p_2 + k_\perp,$
the 
  $d^4k= \frac{s}{2} \, d \alpha_k \, d\beta_k \, d^2k_\perp$ integration of   Fig.\ref{FigkT} reduces at large $s$ in a 2-d integration, when setting $\alpha_k\simeq0$ ($\beta_k\simeq0$) in the upper  (resp. lower) blob and integrating over $\beta_k$ (resp. $\alpha_k$). The tensor connecting upper and lower blob simplifies since only {\it non-sense} gluon polarizations propagates (along $p_1$ ($p_2$)
in  upper (lower) blob) at large $s$.
This results into the  representation (involving impact factors ${\cal J}$)
$$
{\cal M} = is\!\int\!\frac{d^2\,\kb}{(2\pi)^4\kb^2\,(\rb -\kb)^2}
{\cal J}^{\gamma^* \to \gamma^*}(\kb,\rb -\kb)\;
{\cal J}^{\gamma^* \to \gamma^*}(-\kb,-\rb +\kb)\,.
$$

\subsection{LL BFKL Pomeron: limitations}
First, at LL the scale $s_0$ entering in the $Y\!\!=\!\ln s/s_0$ 
resummation is not fixed. Running and scale  

\begin{wrapfigure}{r}{0.58\columnwidth}
\vspace{-.1cm}\hspace{-.4cm}
\begin{tabular}{ccc}
\epsfig{file=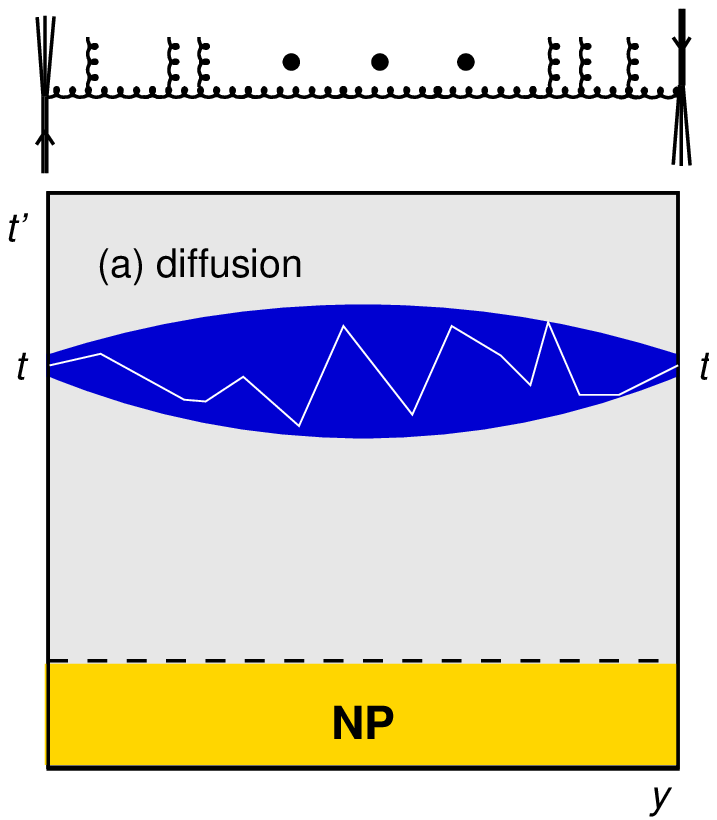,width=\widttt}\hspace{-.2cm}
&
\epsfig{file=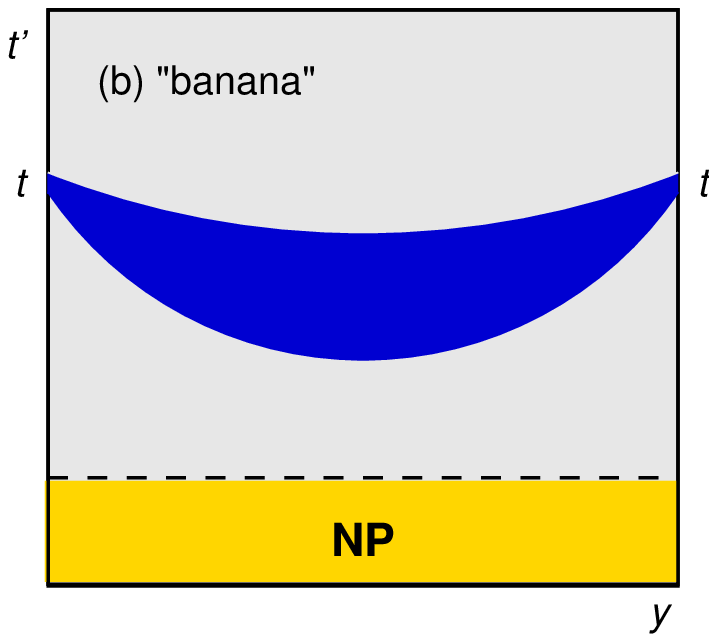,width=\widttt}\hspace{-.2cm}
&\epsfig{file=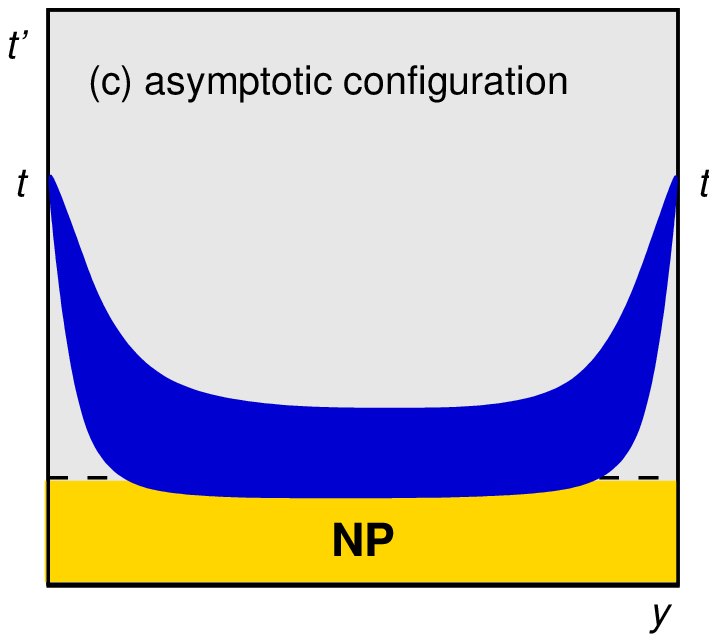,width=\widttt}
\end{tabular}\vspace{-.3cm}
\caption{Diffusion along the BFKL ladder.}
\label{FigCigar}\vspace{-.4cm}
\end{wrapfigure}
\noi   
fixing of $\alpha_S$ are not prescribed at LL. Second, 
energy-momentum is not conserved in the BFKL approach (this remains at any order:
NLL, NNLL, ...), but	 naturally implemented 
(vanishing  of  the first moment of the splitting functions)
in the usual collinear renormalisation group approach (\`a la  DGLAP\cite{dglap}),
since 
one starts with non local
matrix elements. The   energy-momentum 
tensor corresponds to their  first moment, 
 protected by radiative corrections. 
Third, diffusion along the 
ladder spoils the IR  safeness of the BFKL Pomeron:    
  	at fixed $\alpha_S$, there is a gaussian  diffusion of $k_T,$   with  a cigar-like picture \cite{BartelsLotter}. The more $s$ increases, the larger is the broadness.
Setting 
$t \!=\!\ln Q^2/\Lambda_{QCD}^2$ 
(fixed from the probes)
 and 
$t '\!=\!\ln k^2/\Lambda_{QCD}^2$
($k^2\sim -k_T^2$ = virtuality of
an  arbitrary exchanged   gluon along the chain),  
 the typical  width of the  cigar is  
$ \Delta t' \sim \sqrt{\alpha_S Y}$ (Fig.\ref{FigCigar}a). 
The Non-Perturbative   domain  is touched when $\Delta t' \! \sim \! \sqrt{\alpha_S Y} \sim t.$
In a simple running   implementation,  the border of the cigar touches NP for $Y \sim b t^3$ ($b=11/12$)
while the center of the  cigar  approaches NP when $Y \sim b t^2$
("banana structure" of Fig.\ref{FigCigar}b).
A more involved treatment of LL BFKL with running coupling 
\cite{CiafaloniColferaiSalamStasto} showed that 
the cigare is ``swallowed'' by NP in the middle of the
ladder (Fig.\ref{FigCigar}c): one  faces tunneling when $Y \sim t,$ meaning that 
IR safety is doubtful.

\subsection{Higher order corrections}
\label{higherorder}

Higher order corrections to BFKL kernel are known at NLL order ($ \alpha_S \sum_n (\alpha_S \, \ln s)^n$  series) \cite{NLLLipatovFadinCamiciCiafaloni}, now  for arbitrary impact parameter.
Impact factors are known in some cases at NLL
($\gamma^* \to \gamma^*$ at \linebreak $t=0$   \cite{BartelsColferaiGiesekeKyrieleisQiao},
	forward jet production \cite{BartelsColferaiVacca},
$\gamma^* \to \rho$ in forward limit \cite{IvanovKotskyPapa}).
This leads to {\it very large cor-\linebreak rections  with respect to LL}.
The main part of these corrections can be obtained from a physical principle, based on a kinematical constraint along the gluon ladder (which is subleading with respect to LL BFKL)\cite{Kwiecinski}. 
However this
could have nothing to do with  
NLL correction: in principle this constraint would be satisfied when including LL+NLL+NNLL+NNNLL+... .
This constraint is  more related to  improved collinear resummed approaches (see bellow) for which the  vanishing of the first moment of the splitting function is natural.
The above perturbative instabilities require  \break  an improved scheme.
Either one can use a physical motivation to fix the scale of the coupling\footnote{The 
running of the coupling constant should be implemented at NLL, while
scale is fixed starting from NNLL.}: \linebreak
this is the basis of  BLM scheme, applied for the
 $\gamma^* \gamma^* \!\!\to\!\!  X$  total cross-section \cite{BrodskyFadinLipatovKimPivovarov}
and for \linebreak the $\gamma^* \gamma^* \to \rho \rho$ exclusive process \cite{EPSW,IvanovPapa}. 
Or one can use a resummed approach  inspired by compatibility with usual
renormalization group approach
    \cite{SalamCiafaloniColferai}. For example in $\gamma^*(Q_1) \gamma^*(Q_2)\!\to\! X,$
  one  includes {\it both} full DGLAP LL  for
                                     $Q_1 \gg Q_2$ 
                                    and anti-DGLAP LL
                                     $Q_1 \ll Q_2,$ 
 fixes the relation between 
 $Y$ and $s$ in a symmetric
           way compatible with DGLAP  and
       implement running\nopagebreak[4] of $\alpha_S.$   
        Coming back to the IR diffusion problem, this scheme enlarges the validity of perturbative\nopagebreak[4] QCD.
A
simplified version \cite{KhozeMartinRyskinStirling} at fixed $\alpha_S$
 results in performing
      in the LL BFKL Green function 
$$\frac{1}{\kb^3 \kb'^3} \int \frac{d \omega}{2 \pi i} \int \frac{d\gamma}{2
  \pi i}
\left(\frac{\kb^2}{\kb'^2}\right)^{\gamma-1/2} \frac{e^{\omega Y}}{\omega-\omega(\gamma)}$$
\nopagebreak[4] the replacement 
$\omega - \omega(\gamma) \!\to\!\omega
  -\omega(\gamma,\omega).$
The $\omega$ integration is performed through contour \pagebreak closing around  the pole at 
 $\omega=\omega(\gamma,\omega),$ and the  
$\gamma$ integration is made using the saddle point approximation at 
large $Y.$ This  
takes into account the main NLL corrections (within 7 \% accuracy).
          	
\subsection{Non-linear regime and saturation}

The Froissart bound should be satisfied at asymptotically large $s$ and for
  each impact parameter $b,$  amplitudes should fulfil $T(s,b)<1$.
Various unitarization and saturation models have been developped.
First, 
the {\it Generalized Leading Log Approximation}, taking into account any {\it fixed} number $n$ of
$t$-channel exchanged reggeons, leads to the 
Bartels, Jaroszewicz, Kwiecinski, Praszalowicz equation \cite{BJKP},  a 2-dimensional quantum mechanical problem
(time $\sim \ln s$) with $n$ sites. 
It i\break s an {\it integrable model} in the large $N_c$ limit \cite{XXXLipatovFaddeevKorchemsky}, the  XXX Heisenberg spin chain (its {\it non-compact} symmetry group $SL(2,C)$ makes the solution non-trivial).
Solution of  BJKP (i.e. energy spectrum $\Rightarrow intercept$) exists for arbitrary $n,$
describing both Pomeron $P\!=\!C=\!+1$ and Odderon $P\!=\!C=\!-1$ exchanges.
For Odderon, $\alpha_O <1$\cite{JanikWosiekKorchemskyKotanskiManashovLipatovdeVega}       
 but it  decouples from Born impact factors.
 A {\it critical} solution ($\alpha_O =1$)   coupled to Born impact
	       factors
            can be obtained
              either from the perturbative Regge approach
 \cite{BartelsLipatovVacca} 
             or from the dipole model \cite{KovchegovSzymanowskiSW}.
Second, the  {\it Extended Generalized Leading Log Approximation}
      \cite{EGLLA}, in which the number of reggeon in
       $t-$channel is non conserved, satisfies full unitarity (in all sub-channel) and is an
{\it effective 2-d field theory} realizing the Gribov  idea of Reggeon field theory in QCD.
Its simplest version, leading to the  Balitski-Kovchegov equation \cite{B-jimwlk,K}, involves
{\it fan-diagrams} (with singlet sub-channels). Loops (in terms of Pomerons) corrections are unknown, and obtaining them would be a major step. Another effective field theory approach has been developped separately \cite{LipatovKirschnerLipatovSzymanowski}.
Precise relationships between effective approaches remains to be clarified. 
	Third, the
multipomeron approach makes contact with  AGK cutting rules
  of pre-QCD \cite{AGKQCD}.
In the large $N_c$ limit, this is the dominant contribution when coupling to
Born impact factors (leading with respect to BJKP), and it leads to
unitarization. 
 Fourth, during the last decade, the 
Color Glass Condensate\cite{b-JIMWLK} and  B-JIMWLK equation were elaborated.
This effective field theory is  based on the scattering picture of a probe off the field of a source, which is treated through a renormalisation group equation
with respect to a longitudinal scale, with an explicit integration out of modes
below this scale. The approach of 
	Balitski\cite{B-jimwlk} relies on the 
scattering of Wilson  loops and computation of interaction of one loop with the field of the other (related to the eikonal phase approach \`a la Nachtmann).
The mean field approximation of the  B-JIMWLK equation
leads to the BK equation.
There is at the moment no clear one-to-one correspondence between EGLLA and CGC, except in the peculiar BK limiting case.
Loops (in terms of Pomerons) corrections are also unknown.
Toy models in 1+0 dimensions are under developpement (Reggeon field theory) to understand these corrections.
Very interesting links exist between saturation models and statistical physics (reaction-diffusion models of the  FKPP class)\cite{PeschanskiMunier-IancuMuellerMunier}. 
These  models provide a saturation scale $Q_s(Y)$  growing with $Y$:
above this scale the scattering amplitude $T$ is small (color transparency),
	and below \nopagebreak[4] it saturates.
This reduces the contribution of gluons with $k^2 < Q_s^2$ and may solve the IR diffusion\nopagebreak[4] problem.

\subsection{Onium-onium scattering as a gold plated experiment: $\gamma^{(*)} \gamma^{(*)}$ at  colliders}

Tests of perturbative QCD  in the Regge limit require  observables which are free of 
IR divergencies,  by
 selecting external or internal probes with  transverse sizes $\ll 1/\Lambda_{QCD}$ 
({\it hard} $\gamma^{*}$,
{\it heavy} meson  ($J/\Psi$,  $\Upsilon$),
		{\it energetic} forward jets) or by 
		choosing large $t.$ 
They should be governed by 
   the \break {\it "soft"}  perturbative dynamics of  QCD (BFKL)
and {\it not} by its {\it collinear} dynamics (DGLAP\cite{dglap}, ERBL\cite{ERBL}):
probes should have comparable transverse sizes.
They should 
allow control  of $k_T$ spreading, that is the transition from linear to non-linear (saturated regime), meaning the possibil- \linebreak ity  of 
varying $s$ for fixed transverse size of the probes.
It should  give access both to forward (i.e. inclusive) and non-forward (i.e. exclusive processes) dynamics,
both testing linear and non-linear regimes.
 $\gamma^{(*)} \gamma^{(*)}$ scattering satisfies all these requirements.

\section{Inclusive and Exclusive tests of  BFKL dynamics}
 \subsection{Hadron-hadron colliders}
\subsubsection*{Mueller-Navelet jets}

\psfrag{De}[cc][cc]{\scalebox{.62}{$\hspace{1.5cm}\begin{array}{l}\Delta \eta =\\
\hspace{-.2cm}\ln\left(\frac{x_1 x_2 s}{k_1 k_2}\right)\end{array}$}}
\psfrag{up}[cc][cc]{\scalebox{.62}{$\hspace{1.2cm} k_1, \, y_1 = \ln\left(\frac{x_1\sqrt{S}}{k_1}\right)$}}
\psfrag{do}[cc][cc]{\scalebox{.62}{\raisebox{-.8\totalheight}{$\hspace{1.2cm} k_2, \, y_2 = \ln\left(\frac{x_2\sqrt{S}}{k_2}\right)$}}}
\psfrag{xu}[cc][cc]{\scalebox{.62}{\raisebox{-2\totalheight}{$\!\!\!\!\!\! x_1$}}}
\psfrag{xd}[cc][cc]{\scalebox{.62}{$\!\!\!\! x_2$}}
\vspace{0cm}
\begin{wrapfigure}{r}{0.3\columnwidth}
\vspace{-3.8cm}
\begin{center}
\hspace{-1.2cm}
\vspace{-.5cm}
\epsfig{width=0.26 \columnwidth, file=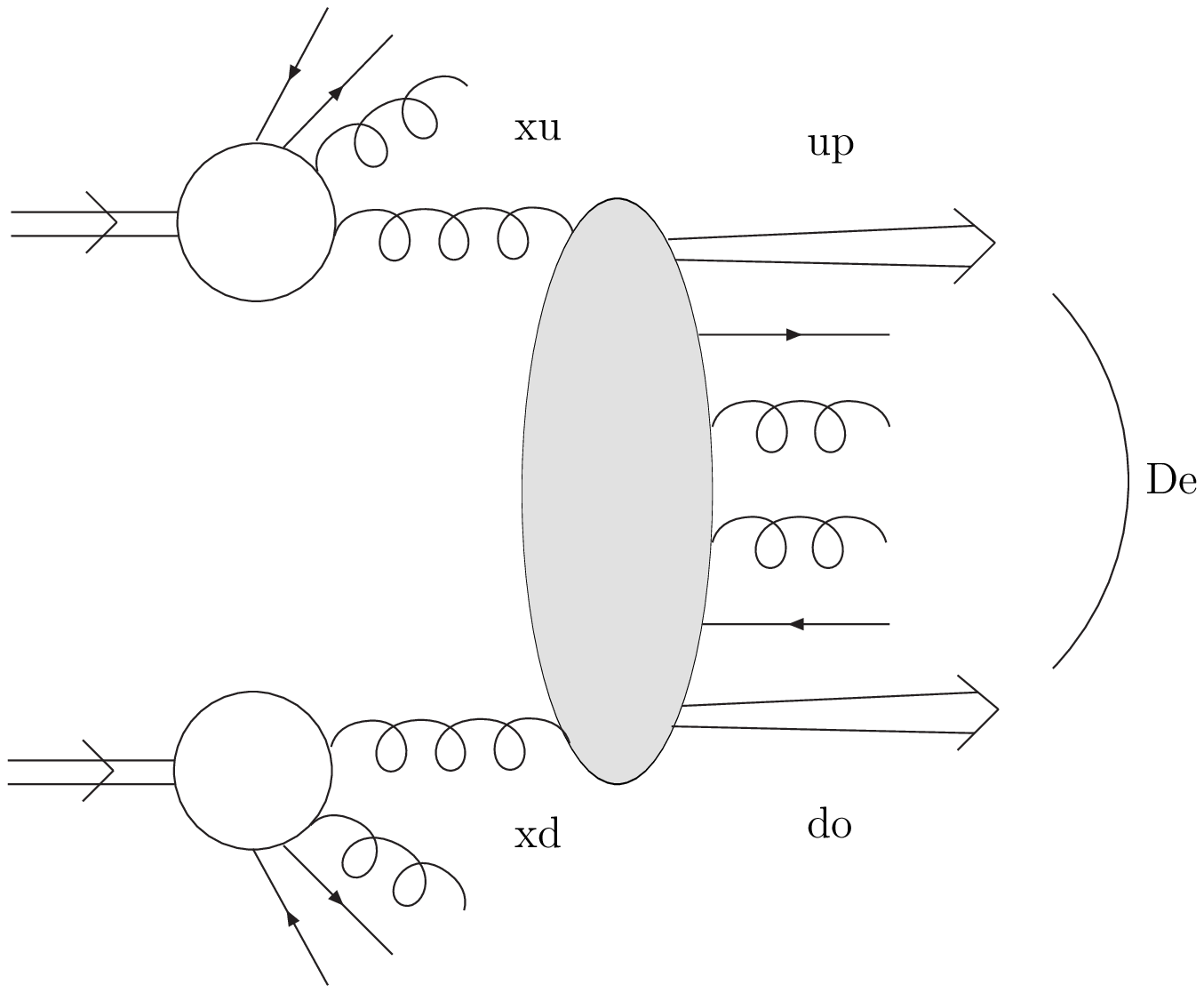}
\end{center}\vspace{-.3cm}
\caption{Mueller-Navelet jets.}
\label{FigMuellerNavelet}\vspace{-.5cm}
\end{wrapfigure}
This test of BFKL at $t=0$ is based on 
the measure for two jets at large $p_T$ (hard scale) separated by a large rapidity $\Delta \eta$
(Fig.\ref{FigMuellerNavelet}). The signal is a decorrelation of relative azimutal angle between emitted jets when increasing  $\Delta \eta.$
Studies were made at LL \cite{MuellerNavelet}, NLL  \cite{MNVerraSchwennsen} 
and resummed NLL \cite{MarquetRoyon}.  Tevatron I data\cite{dataTevatron1} agreed \cite{tevatron1} with the modified BFKL approach \cite{Kwiecinski} (see section \ref{higherorder}).
	The  
measurement should be performed soon at  CDF for $\Delta \eta$ up to 12, and
presumably at LHC. 
	

\vn
\subsubsection*{Diffractive high energy double jet production}

\begin{wrapfigure}{r}{0.25\columnwidth}
\vspace{-2.2cm}
\begin{center}
\hspace{-.8cm}
\epsfig{width=0.15 \columnwidth, file=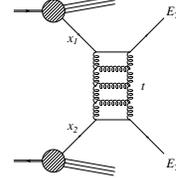}
\end{center}\vspace{-.6cm}
\caption{Diffractive high energy jet production.}
\label{FigDiffJet}\vspace{-.3cm}
\end{wrapfigure}
The idea is to measure  two jets with a gap in rapidity (Fig.\ref{FigDiffJet}), with hard  scales provided 
by the energies $E_T$ of the jets \cite{Mueller-Tang}. This tests BFKL  at $t \neq 0.$ Taking into account non perturbative gap survival rapidity\cite{EnbergIngelmanMotyka}, one can correctly describe the Tevatron data
\cite{DataET}.

\vn
\subsubsection*{High $p_T$ jet production} 

\psfrag{g}[cc][cc]{\scalebox{.6}{$\gamma$}}
\begin{wrapfigure}{r}{0.28\columnwidth}
\vspace{-1.3cm}
\begin{center}
\hspace{.5cm}
\epsfig{width=0.25 \columnwidth, file=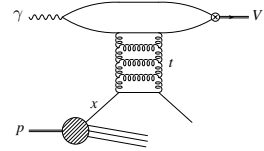}
\end{center}\vspace{-.65cm}
\caption{Exclusive vector meson production at large $t$.}
\label{FigVectorMeson}\vspace{-.5cm}
\end{wrapfigure}
This has been studied at LL and NLL \cite{highpT}. It  
relies 
on computation of impact factors, kernel and Green function at LL and NLL order. The effective jet vertex requires a precise definition of the emitted jet (made of one or two $s-$channel
emitted particle   at NLL), and
modeling of proton impact factor 
(the only hard 
 scale is  $p_T^2$).

\subsection{HERA}


\subsubsection*{DIS and diffractive DIS}
\psfrag{r}[ll][ll]{\scalebox{.35}{$\rho\,\,$}}
\begin{wrapfigure}{r}{0.27\columnwidth}\vspace{-2.4cm}
\begin{center}
\hspace{-.3cm}\epsfig{file=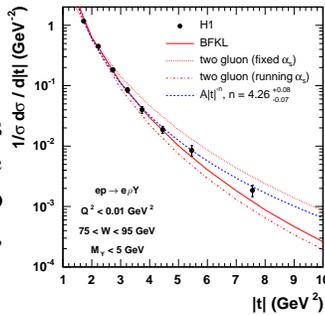,width=0.29\columnwidth}
\end{center}\vspace{-.8cm}
\caption{Exclusive vector meson production at HERA.}
\label{FigVectorMesonData}\vspace{-.7cm}
\end{wrapfigure}

 $Q^2$ being the only hard scale for DIS,  a model for the proton
 is needed \cite{PeschanskiNaveletRoyonSW-GolecBiernatKwiecinski}.
 BFKL  (at $t=0$) and DGLAP (NLL) both describe the data\cite{DISH1ZEUS}.
 Diffractive DIS\cite{ingelman}, corresponding experimentally to\nopagebreak[4] a gap in the detector between the proton remants and the jets\cite{dataDiffractionDIS}, can\nopagebreak[4] be described both within collinear and
BFKL approaches\cite{diffractionModels}.
\vn
\subsubsection*{Energetic forward jet and $\pi^0$ production}

 It is a test of
BFKL at  $t=0,$ with hard scales  given by the $\gamma^*$ virtuality and the jet energy
 \cite{Mueller-BartelsLoeweDeRoeck-KwiecinskiMartinSutton-BartelsDelDucaWüsthoff}. Data\cite{H1ZEUSforward}  favor BFKL but cannot exclude a partonic scenario\cite{fontannaz}.\break
\vspace{-.95cm}
\subsubsection*{Exclusive vector meson production at large $t$}
This test of BFKL at large $t,$ which provides the hard scale (Fig.\ref{FigVectorMeson})\pagebreak \cite{ForshawRyskin-BartelsForshawLotterWüsthoff-ForshawMotykaEnbergPoludniowski}, was made  for H1, ZEUS data and 
favor BFKL (Fig.\ref{FigVectorMesonData}).
Problems with data  remains for the  spin density matrix.

\subsection{$\gamma^* \gamma^*$ at LEP2}

\begin{wrapfigure}{r}{0.75\columnwidth}
\vspace{-2.4cm}
\begin{center}
\begin{tabular}{ccc}
\hspace{-.5cm}\raisebox{-.04 \totalheight}{\epsfig{file=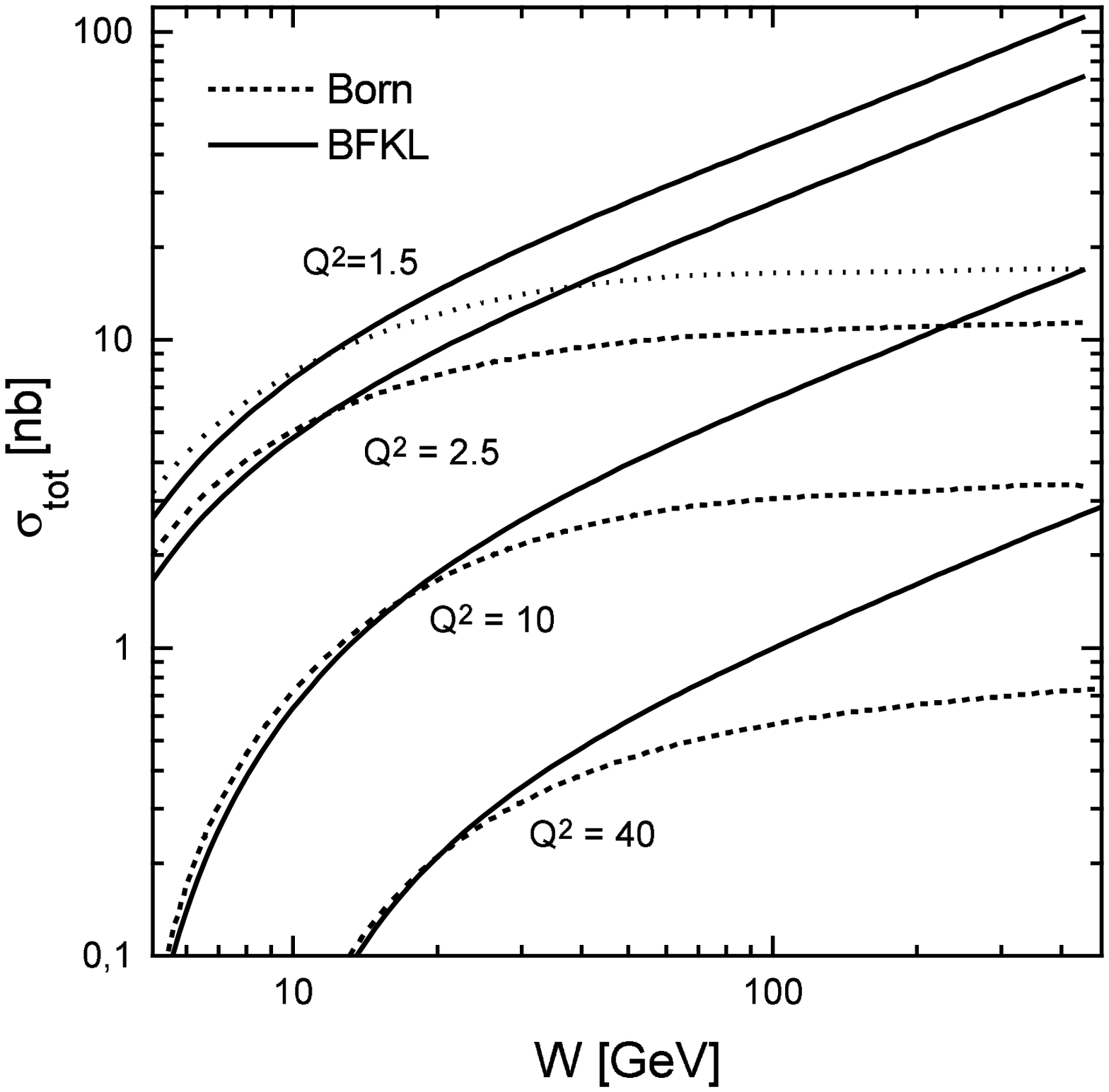,width=0.265\columnwidth}}& \hspace{-.6cm}\epsfig{file=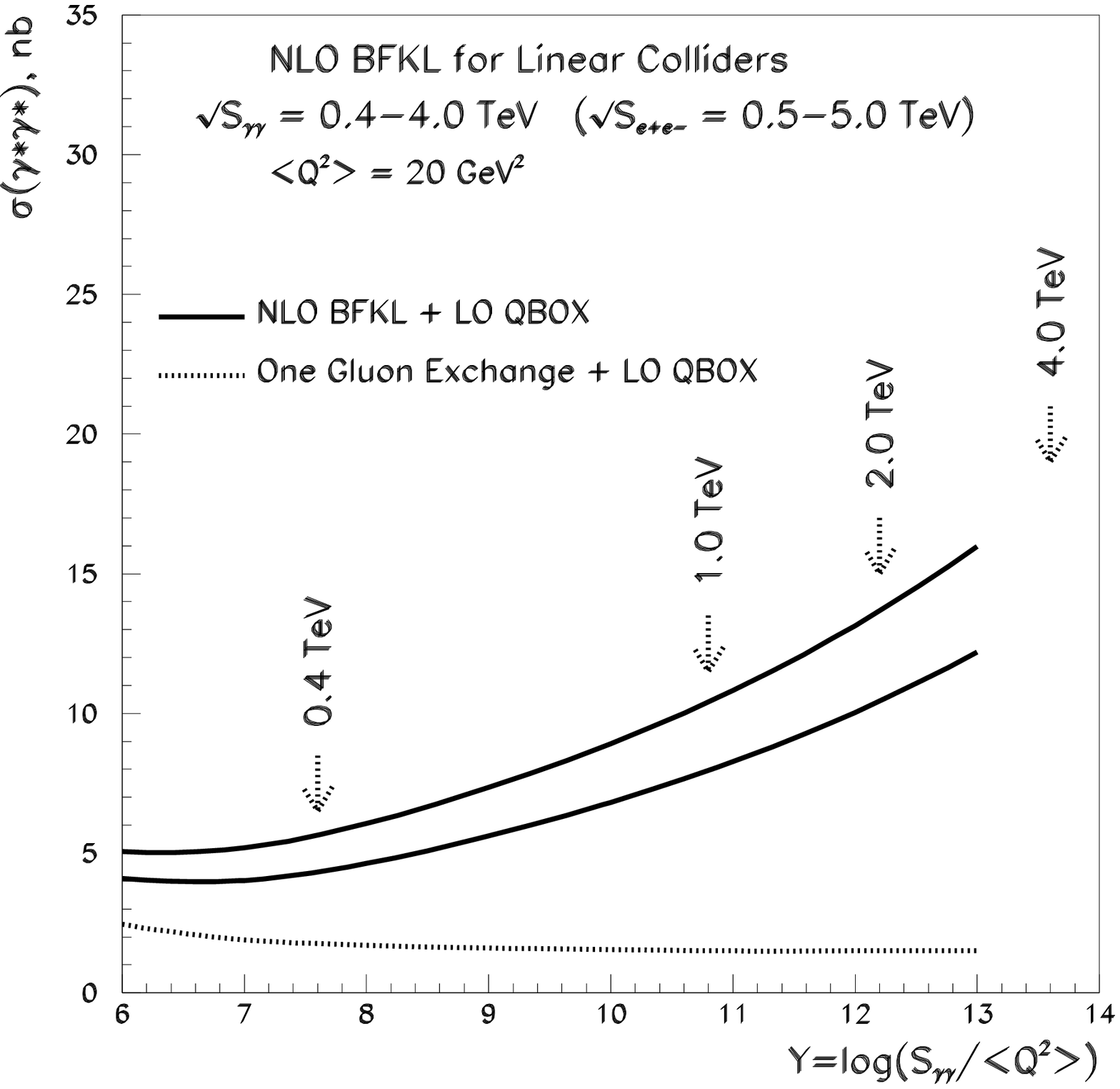,width=0.25\columnwidth} &\hspace{-.7cm}\epsfig{file=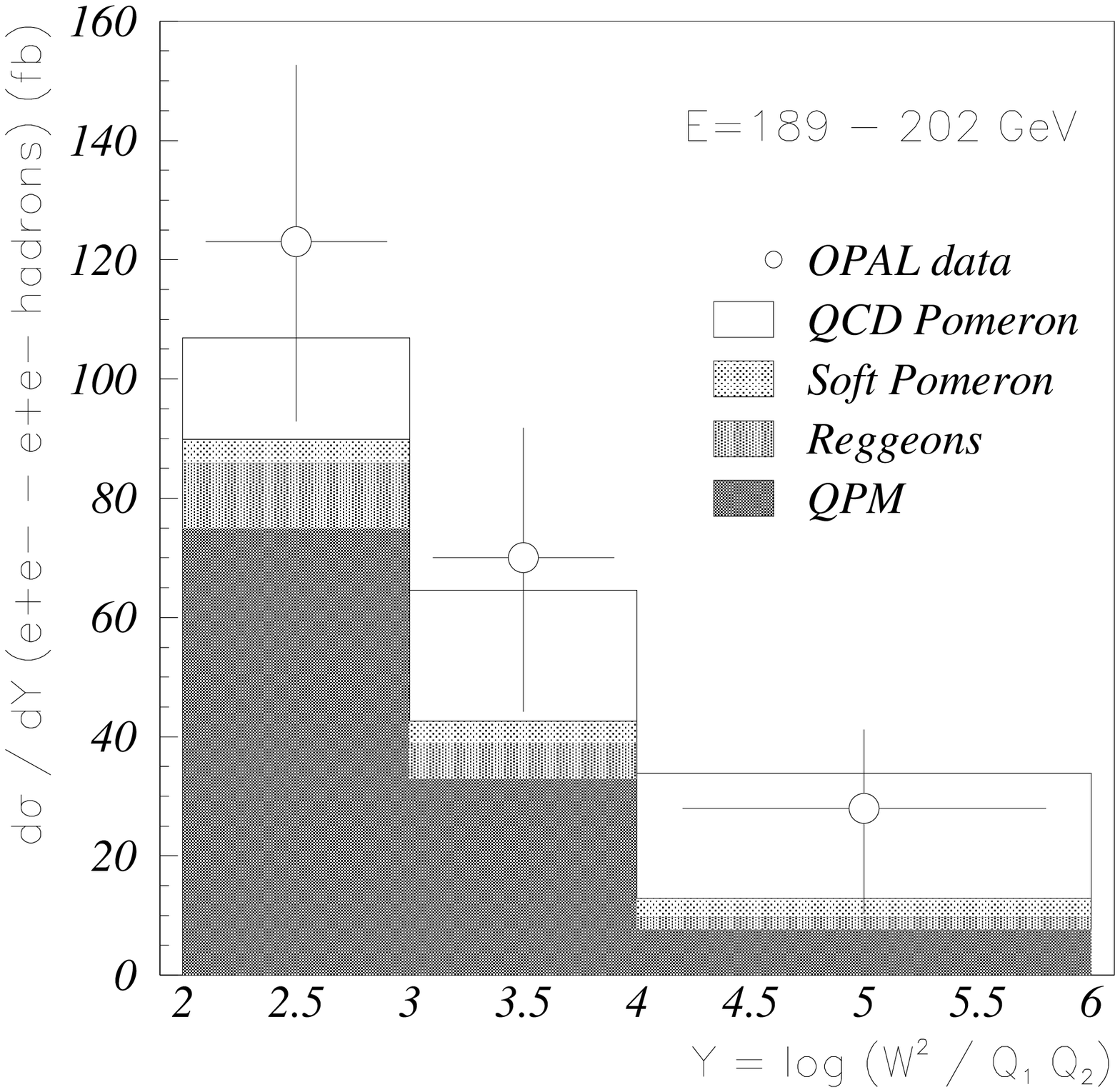,width=0.25\columnwidth}
\end{tabular}
\vspace{-.5cm}
\end{center}
\vspace{-.4cm}
\caption{
Modified LL BFKL (a) and BLM scale-fixed NLL BFKL (b) predictions versus Born.  OPAL data versus modified LL 
(c).
}
\label{FigTotalXS}\vspace{-.4cm}
\end{wrapfigure}
The   LEP2 available energy ($\sqrt{s_{e^+e^-}}$ = 183 to 202 GeV) allowed tests of the total $\gamma^* \gamma^*$ cross-section. 
This process was studied with LL BFKL\cite{BartelsDeRoeckLotter-BrodskyHautmannSoper}, dipole model\cite{BoonekampDeRoeckRoyonSW,BialasCzyzFlorkowski}, modified LL BFKL (based on kinematical constraints)\cite{Kwiecinski-Motyka} (Fig.\ref{FigTotalXS}a), NLL BFKL\cite{BrodskyFadinLipatovKimPivovarov} (Fig.\ref{FigTotalXS}b).
Fig.\ref{FigTotalXS}c  displays the comparison\cite{Kwiecinski-Motyka} between 
 modified LL BFKL, 
including quark box (simulating usual DGLAP for $Q_1 \sim Q_2$), soft Pomeron  and reggeon contributions,
and  OPAL data.
%
 	Born 2 gluon exchange and  quark exchange are too small in the large $Y$ set of the data.
	LL BFKL is too high (even including quark mass effects\cite{BartelsEwerzStaritzbichler}).
	Scenarios with modified BFKL or NLL BFKL with  BLM scale fixing were plausible,
but lack of statistics\cite{dataLEP} (minimal detection angle of only $30$ mrad, 
luminosity and energy  limited) forbade any 
conclusion.\break

\vspace{-.5cm}
\section{Onium-onium scattering  at ILC collider}
\subsection{Sources of photons}

The direct $\gamma \gamma$ cross-section (box diagram) is out of reach experimentally. For example,
$\sigma^{\gamma\gamma \to \gamma \gamma} \sim 10^{-64} (\omega_\gamma / {\rm eV})^6 {\rm cm^2},$
that is $10^{-65}{\rm cm^2}$
for visible light ($\omega\! \sim \!1$ eV)!
Photons can be produced either, 
using the Fermi, Weizsäcker, Williams  idea that the field of a charged particle is a  flux of equivalent photon (which are almost real), from a high luminosity collider 
($A p,$ $pp$, $e^+p$, $e^+ e^-$) or from Compton backscattering to pump the energy of electrons of a storage ring or of a collider.
\vspace{-.5cm}\noi
\subsubsection*{Photon colliders: hadron and nucleus colliders}
To produce high energy $\omega=z E_{Ze}$ photons with high luminosity, the equivalent photon approximation
$$P_{\gamma/Ze}(z,Q^2) \sim Z^2 \, \alpha_{em} /{(z \, Q^2)}$$
implies that  one can use either a high energy (to compensate the  $1/z$ pole) and high luminosity hadron collider (LHC, Tevatron),
or a  heavy nucleus collider ($Z^2$ then balance the lower luminosity) (RHIC,  LHC) \footnote{see Nystrand's talk}. 
At  LHC, both modes would give comparable fluxes of photons.
However,  $\gamma\gamma$ events are poluted by pure (soft) hadronic interactions between source of photons, since hadrons or nucleus
are sensitive to strong interaction.
   One needs to select peculiar ultraperipheral events
for which the typical impact parameter $b$ between hadrons (nucleus) exceeds $1/\Lambda_{QCD}.$
This is possible experimentally with very forward detectors, with
(anti)tagging protons:
		forward detector at CDF (with coming data), 
 LHC detectors (Roman pots) suggested at 420 m (FP420 at CMS and ATLAS) and 220 m 
 (RP200 at ATLAS) from the Interaction Point at LHC. These last detectors 
		are   very promising  for both 
$\gamma \gamma$ and  hadronic diffractive physics  (ex:
Higgs exclusive production, MSSM, QCD), but they suffer 
from non trivial problems with fast time trigger (long  distance
from IP to the detector to be comparared with the rate of events at high luminosity).
Combining both detectors would increase acceptance.			
Note that $b$ is not directly reconstructed, and that 
 survival probability have to be taken into account (non-perturbative ingredient).
The above situation should be contrasted with processes involving $e^\pm,$ which are  not directly affected by strong interaction.
This is the key reason why $e^+ e^-$ {\it colliders are the cleanest solution in principle for
       $\gamma^{(*)}\gamma^{(*)}$ physics}, both from a theoretical  and from an experimental point of view.

\vn
\subsubsection*{Photon colliders: $e \to \gamma$ conversion}

At  $e^+e^-$ colliders, 
a small number  of 
photons, of soft spectrum ($dn_\gamma \!\sim \!0.03 \, d \omega/\omega$), is produced: 
\beqd
L_{\gamma \gamma} (W_{\gamma}/(2 E_e)>0.1) \!\sim \!10^{-2} \, L_{e^+e^-} \quad {\rm and} \quad
L_{\gamma \gamma} (W_{\gamma}/(2 E_e)>0.5) \!\sim \!0.4 \, 10^{-3} \, L_{e^+e^-}\,. 
\vspace{-.05cm}
\eeqd
\begin{wrapfigure}{r}{0.198\columnwidth}\vspace{-1.1cm}
\begin{center}
\epsfig{file=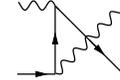,width=0.1\columnwidth}
\end{center}\vspace{-.75cm}
\caption{$u$-channel diagram for Compton scattering.}
\label{FigComptonU}\vspace{-.5cm}
\end{wrapfigure}
To produce a photon collider, the Novosibirsk group 
suggested\cite{Novosibirsk} to reconsider the
use of Compton backscattering of a laser on the high energy electron beam of a collider\cite{backscat}. Due to the $u$-channel diagram of Fig.\ref{FigComptonU}, which has an almost vanishing
propagator, the cross-section is peaked in the backward direction. In this  direction, almost all the energy of the incoming electron is transfered to the outgoing photon  (up to 82 \% at ILC 500 GeV).   The limit comes from the fact that one does not want to    reconvert $\gamma$ in $e^+e^-$ pairs!). The corresponding number of equivalent photons is of the order of 1 if the beam has a small size, with laser flash energy of $1-10$ J. The photon beam follows the direction of the incoming electron beam with an opening angle of $1/\gamma_e.$ Due to the very  good focussing of electrons beams expected at ILC, this  is \break the main effect  limiting the luminosity  in $\gamma$  mode: the distance 
\vspace{-.1cm}

\begin{wrapfigure}{r}{0.33\columnwidth}\vspace{-1.4cm}
\begin{center}
\hspace{-.3cm}\epsfig{file=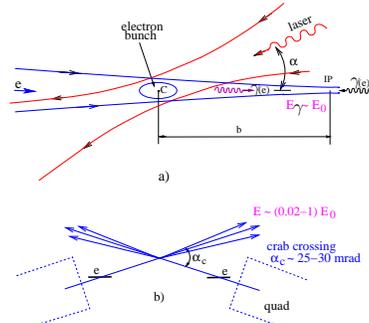,width=0.33\columnwidth}
\end{center}\vspace{-.8cm}
\caption{$\gamma\gamma / \gamma e$ collider with cross-crab angle.}
\label{FigCrossCrab}\vspace{-.3cm}
\end{wrapfigure}
 \noi 
  $b$  between conversion region and Interaction Point
is $\sim 1.5$ mm, making
 impossible to use a magnet to deflect the low energy outgoing electron beam.
 It has been suggested to use a non zero scattering angle between the two incoming beams to remove them (see Fig.\ref{FigCrossCrab}).
In order to compensate the potential lost luminosity
with non zero scattering angle, {\it crab-cross} scattering
is studied (the paquet is not aligned with the direction of its propagation, like a crab).
The luminosity could reach $0.17\,  L_{e^+e^-},$ a very interesting
value
 since the cross-sections in $\gamma\gamma$ are usually one order of magnitude higher than for $e^+e^-.$ The matrix element of the Compton process is  
 \vspace{-.1cm}

\begin{wrapfigure}{r}{0.495\columnwidth}\vspace{-1.4cm}
\begin{center}
\begin{tabular}{cc}
\hspace{-1.1cm}\epsfig{file=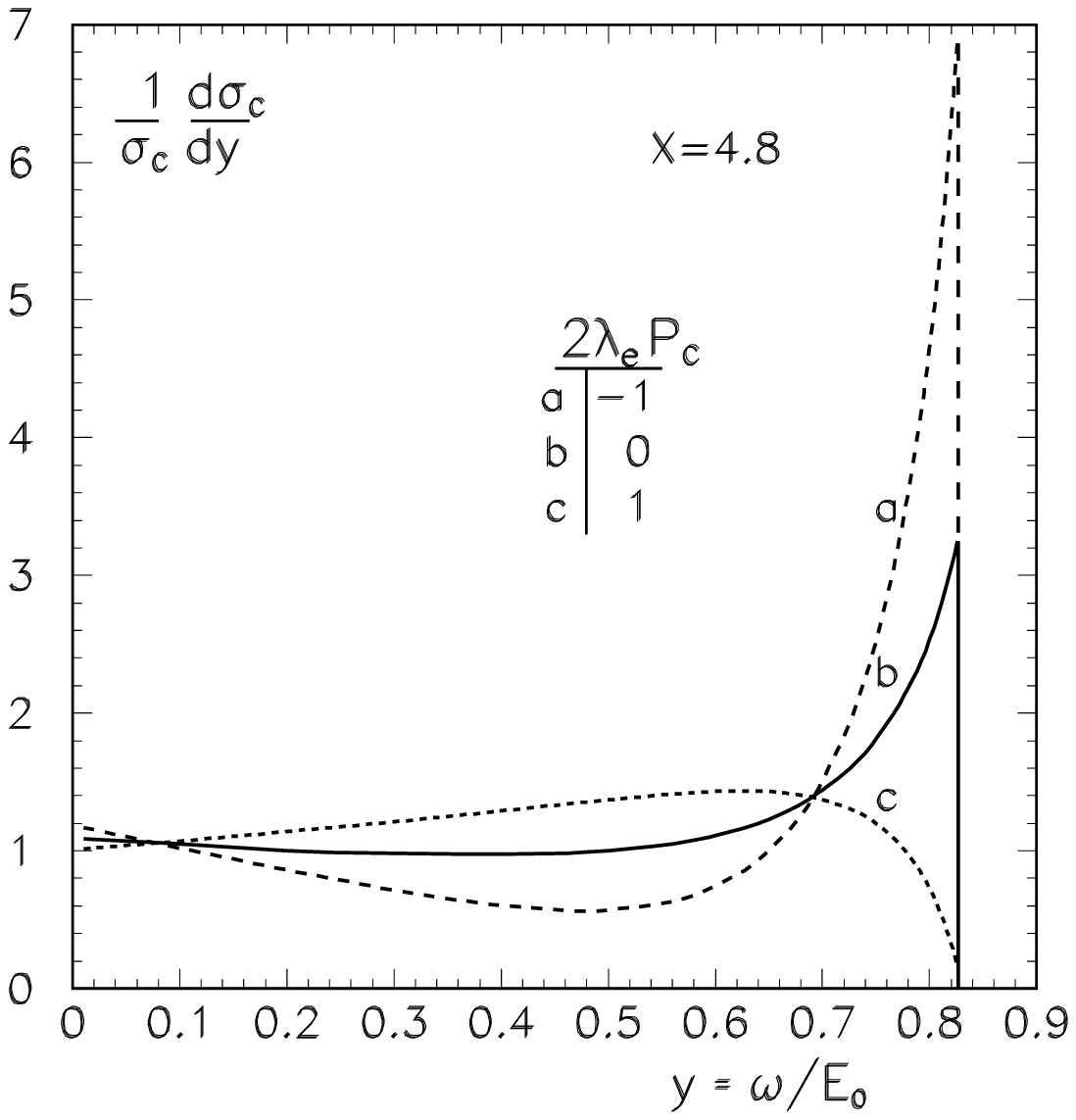,width=0.32\columnwidth} &\hspace{-1.3cm}
\epsfig{file=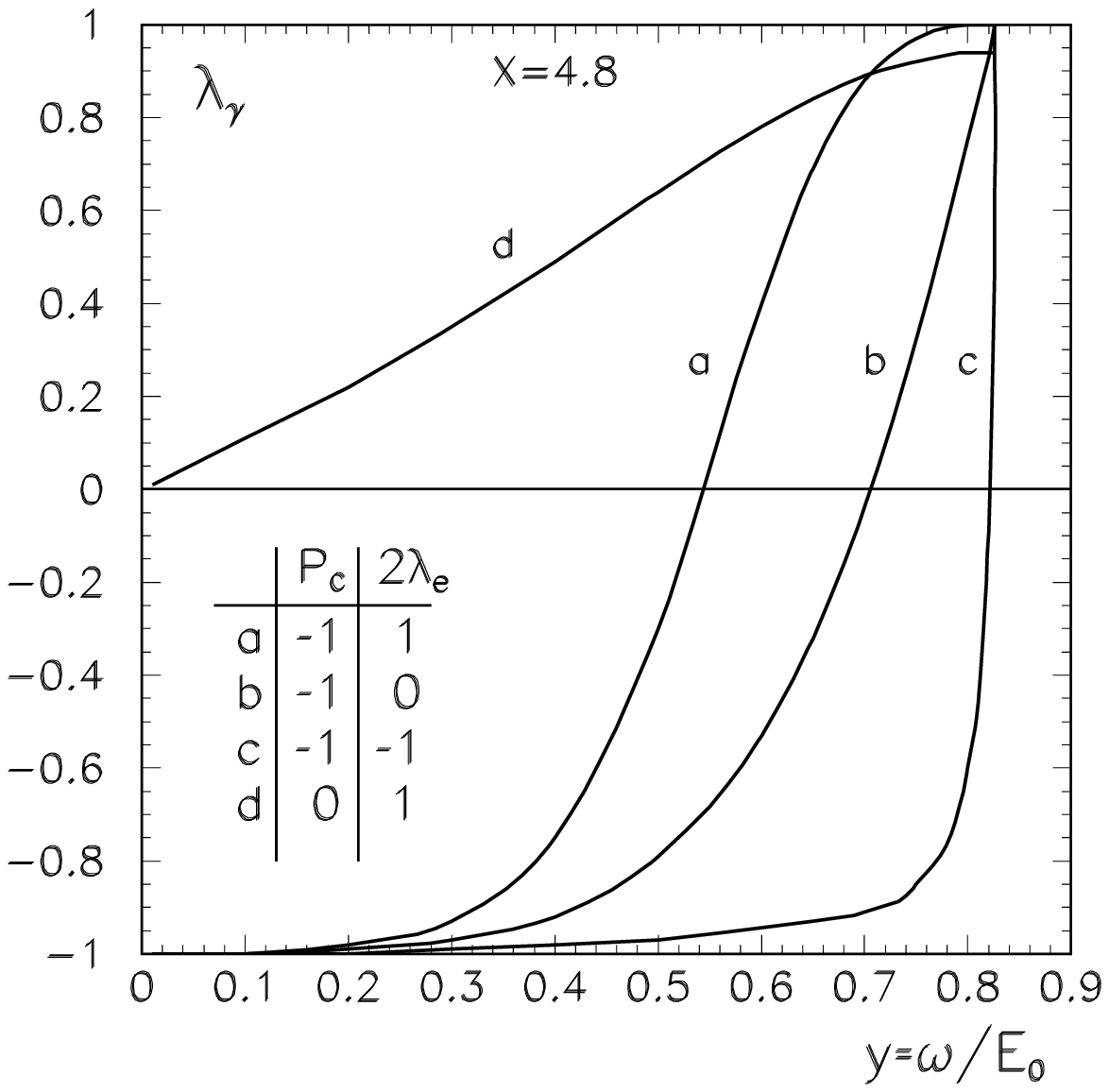,width=0.32\columnwidth}
\end{tabular}
\end{center}\vspace{-.9cm}
\caption{Spectrum (left) and average helicity  (right) of the Compton-scattered photons.}
\label{FigPolGammaGamma}\vspace{-.8cm}
\end{wrapfigure}
\noi   helicity-conserving except for  the term proportional to the electron mass, which is helicity-flip, and {\it dominates in the backward region}. This
 provides a very elegant way of {\it producing quasi monochromatic photons of maximal energy and given polarization,}
 by using $2 \lambda_e P_c=-1$
($\lambda_e$ = mean electron helicity and $P_c$ = mean laser photon circular polarization), see Fig.\ref{FigPolGammaGamma}.
Note that 
WW distribution is sharply peaked around almost on-shell and soft photons:
in $\gamma e$ or $\gamma \gamma$ mode,  
in order to use perturbative QCD, one needs to provide  hard scales,
 from the outgoing state ($J/\Psi$,...) or from large $t$.
Ingoing $\gamma^*$ hard states are  provided only in $e^+e^-$ mode with double tagged outgoing leptons.

\begin{wrapfigure}{r}{0.4\columnwidth}\vspace{-1.8cm}
\begin{center}
\hspace{-.7cm}\epsfig{file=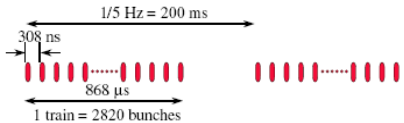,width=0.4\columnwidth} 
\end{center}\vspace{-.9cm}
\caption{Paquet structure for ILC.}
\label{FigPaquet}\vspace{-.6cm}
\end{wrapfigure}

\subsection{ILC project}

The ILC  budget estimate is  6.65 G\$,
comparable to the cost of the LHC  when including  pre-existing facilities.
\vn
\subsubsection*{Reference Design Report for ILC}

\begin{wrapfigure}{r}{0.215\columnwidth}\vspace{-.8cm}
\begin{center}
\vspace{-.1cm}
\hspace{-.4cm}\includegraphics[width=3.4cm,bb=0 0 482 475, clip]{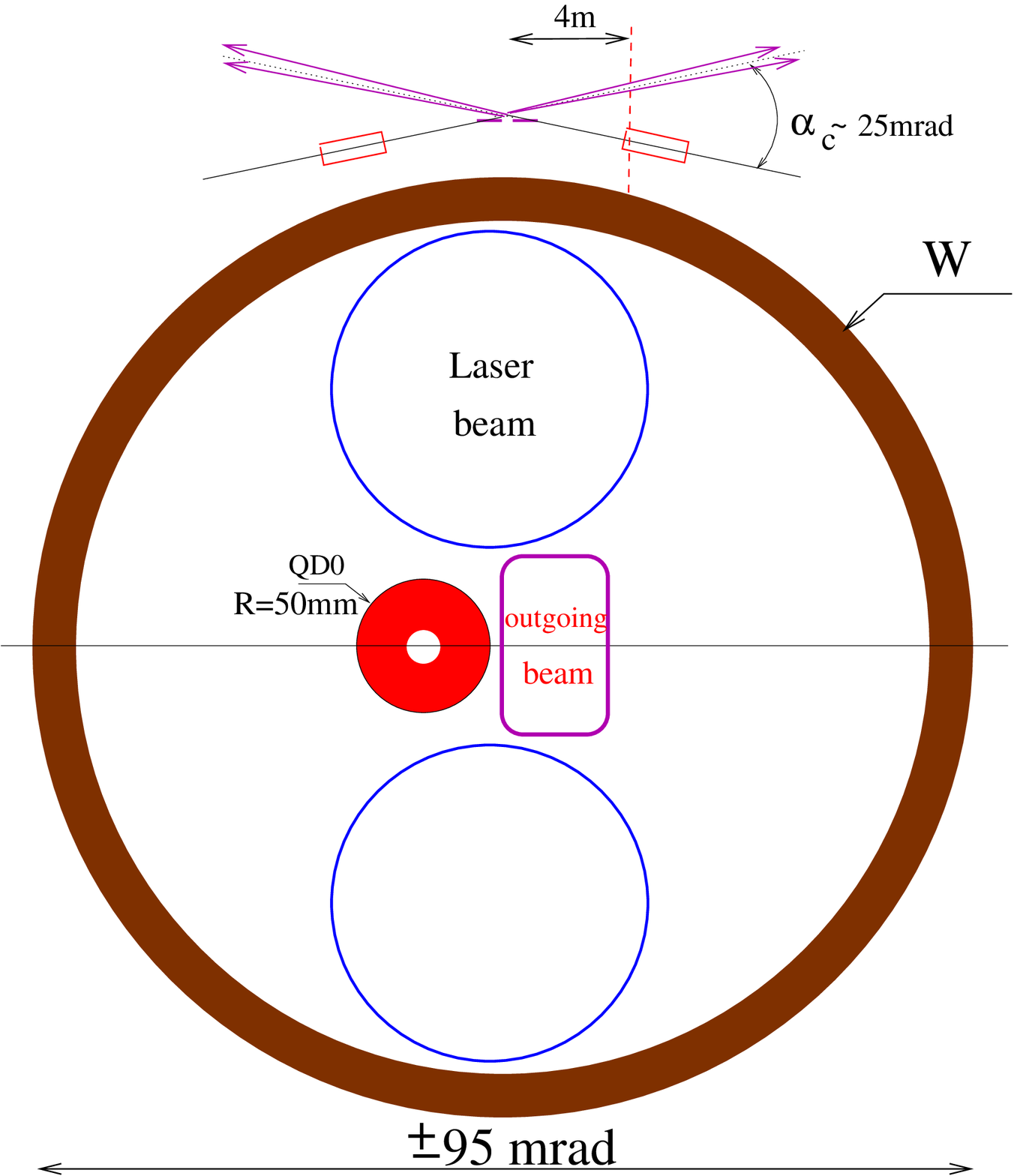}
\end{center}\vspace{-.5cm}
\caption{
Quad, electron and laser beams.
}
\label{FigMirors}\vspace{-.6cm}
\end{wrapfigure}
${\sqrt s_{e^+e^-}}$
should 
be 500 GeV, with a 
 luminosity of  $125 \,{\rm fb^{-1}}$  per year within  4 years of running, 
with a possible scan in energy between $200 \,{\rm GeV}$ and $500 \,{\rm GeV}.$
An upgrade at  $1\,{\rm TeV},$ with a luminosity of $1 \,{\rm ab^{-1}}$  within 3 to 4 years is planned 
(see Fig.\ref{FigPaquet} for the rather  intricate structure needed for the paquets)\cite{ILC}.
There are
 non trivial technological  problem  for extracting the outgoing beam.
At the moment,  3 options are considered  for  the  scattering angle: 2 mrad, 14 mrad and 20 mrad, with in each case a hole  in the detector at that angle to let the outgoing beam get through
toward  the beam dump  (reducing the  acceptance in the forward calorimeter). Crab-cross scattering
is needed to  get high luminosity.
	Two interaction  regions are highly desirable: one which could be at low crossing-angle, and one compatible
 with $e\gamma$ and
$\gamma\gamma$ physics (through single or double laser Compton backscattering).
 $\gamma\gamma$  mode  leads to the severe constraint  that $\alpha_c \!\! >$ 25 mrad \footnote{last quadrupole ($\oslash$ =5cm) at 4m from IP and horizontal disruption angle=12.5 mrad, thus
 0125+5/400=25 mrad.}. The mirors could be placed either inside 
or outside the detector,
depending on the chosen technology,
with {\it almost no space
 for any forward detector in a cone of  95 mrad} (Fig.\ref{FigMirors}).
If the cheaper option suggested by Telnov
(single detector + single interaction point + single extraction line, without displacement of the detector between 2 interaction points) would be chosen,   diffractive physics could become very difficult.

\vn
\subsubsection*{Detectors at ILC}

\begin{wrapfigure}{r}{0.54\columnwidth}\vspace{-1.7cm}
\begin{center}
\begin{tabular}{lc} \hspace{-2.1cm}\epsfxsize=4.5cm{\centerline{\epsfbox{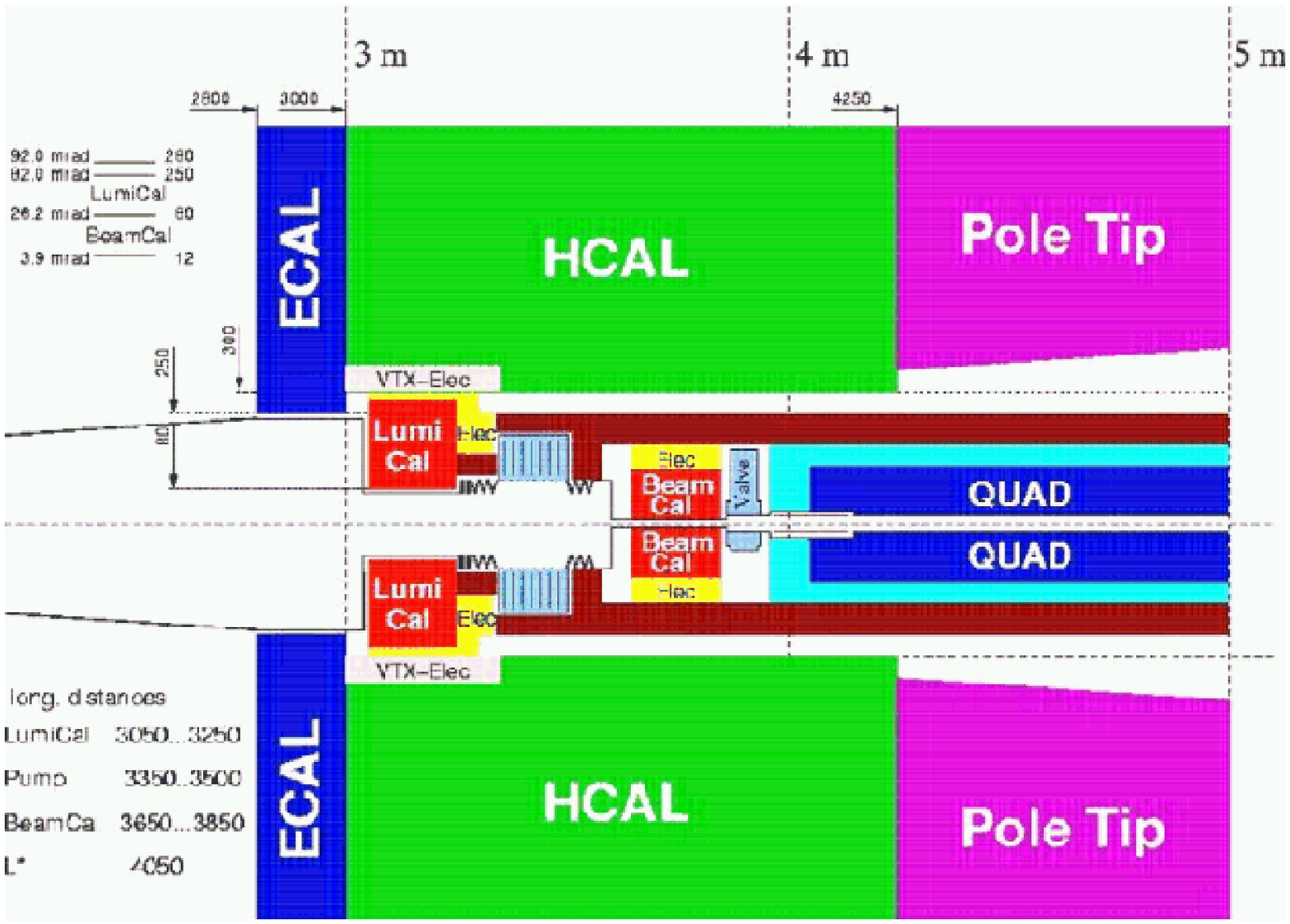}}} & \hspace{-4.25cm}\epsfxsize=3.4cm{\centerline{\epsfbox{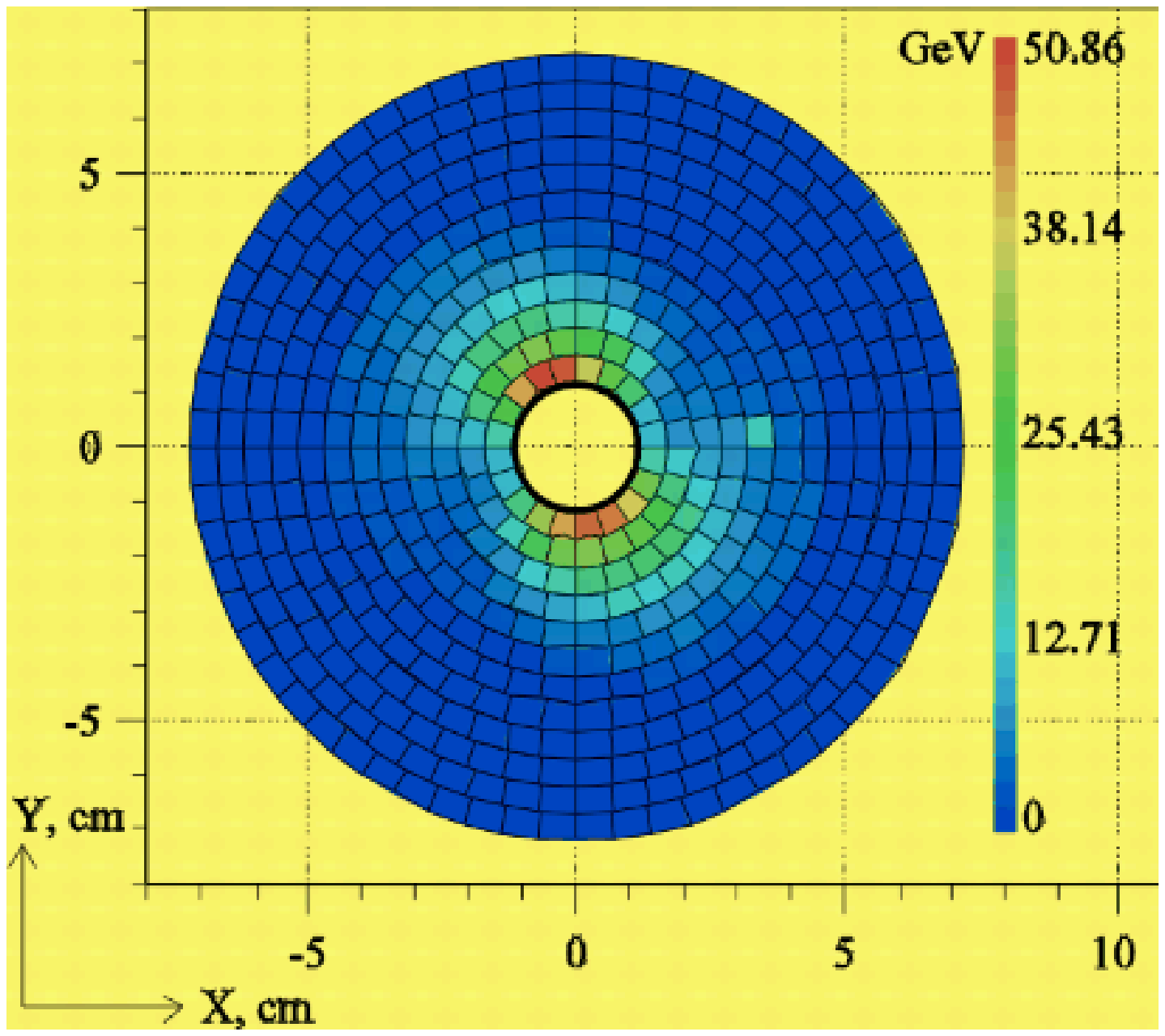}}}
\end{tabular}
\end{center}\vspace{-.7cm}
\caption{LDC  (a). Beamstrahlung in BeamCal (b).}
\label{FigLDC}\vspace{-.4cm}
\end{wrapfigure}

Each of the 4 detector concepts (GLD, LDC, Sid and 4th (sic)) involves a very forward electromagnetic calorimeter for luminosity measurement,
with tagging angle for outgoing leptons down to 5 mrad (10 years ago, 20 mrad was almost impossible!).
	It is ideal  for diffractive physics, which cross-sections are sharply peaked in the very forward region.
	The luminosity is enough to\linebreak  get high statistics, even for exclusive events.
 For example,  LDC (Fig.\ref{FigLDC}a) contains a  BeamCal
at 3.65 m from the vertex\cite{LDC}.
The main background is due 
to beamstrahlung photons, 
 leading
to energy deposit in 
cells close from the beampipe (Fig.\ref{FigLDC}b).
This implies cutting-off the cells for lepton tagging with 
$E_{min}\!\!=\!\!100$ GeV, $\theta_{min}$ = 4 mrad
(and to lower energies for large angles).

\subsection{$\gamma^* \gamma^*  \to hadrons$ total cross-section}

 In comparison to LEP, $s$ would be higher,
 the luminosity would be much higher (a factor $\sim  10^3 $), and 
	detectors would 
give access to events closer to the beampipe (LEP: $\theta_{min} \geq$ 25 to 30\nopagebreak[4] mrad).
One can thus hope to get a much better access to QCD in perturbative Regge limit.
To have enough statistics in order to see a  BFKL enhancement at TESLA, it was considered to be  important to get access down to $\theta_{min} \simeq$ 25 to 20 mrad\cite{BoonekampDeRoeckRoyonSW}. 
	 Probably this could be extended up to 30 mrad due to the expected luminosity
	(a factor 2 to 3  of luminosity higher than TESLA project).  With detection
	 down to 4 mrad, 
this is  thus not anymore a critical parameter\footnote{Note that within a $\gamma e$ and $\gamma \gamma$ option, the Telnov suggestion would forbid any forward detector bellow 100 mrad.}. In a modified LL  BFKL  scenario, one expects around $10^4$ events per year with $\theta_{min} \simeq$ 10 mrad.


\subsection{$\gamma^{(*)} \gamma^{(*)}$ exclusive processes and other QCD studies}
 
In the  $\gamma \gamma$ case ($e^+e^-$ without tagging or  $\gamma \gamma$ collider option), one can consider
any diffractive process of type $\gamma \gamma \to J/\Psi J/\Psi$  \cite{KwiecinskiMotykaJPsi} (or other heavy produced state). The hard scale is provided by  the charmed quark mass, with
an expected number of events for ILC around $9 \,10^4$.
Due to the small detection angle offered by Beamcal, one coud also  investigate the process
$\gamma^* \gamma^* \to \rho_L^0  \;\rho_L^0$\cite{PSW,EPSW,gdatda,SSW,PapaVera} from
$e^+e^- \to e^+e^- \rho_L^0  \;\rho_L^0$ with double tagged out-going leptons\cite{slidestalk1}. The channel $\gamma^* \gamma^*$ is also a gold place for production of  $C$ even resonances, such as $\pi^0,$ $\eta$, $\eta'$, $f_2$. It would be a good place where looking for
the elusive odderon, in  processes like $\gamma^{(*)} \gamma^{(*)} \to \eta_c \, \eta_c$\cite{ewerzOdderon}. Beside the regge limit,  ILC would be also a nice place for finding exotic states like $q \bar{q} g$  with $J^{PC}=1^{-+}$\cite{AnikinPireSzymanowskiSW}. Finally, it could have a great potential for photon structure studies\cite{Klasen}.

\begin{footnotesize}
\bibliographystyle{blois07}

\end{footnotesize}

\end{document}